\documentstyle[twoside,fleqn,espcrc2]{article}

\def\frac#1/#2{\leavevmode\kern.1em
\raise.5ex\hbox{\the\scriptfont0 #1}\kern-.1em
/\kern-.15em\lower.25ex\hbox{\the\scriptfont0 #2}}
\def\bffrac#1/#2{\leavevmode\kern.1em
\raise.5ex\hbox{$\bf\scriptstyle #1$}\kern-.1em
{\bf/}\kern-.15em\lower.25ex\hbox{$\bf\scriptstyle #2$}}

\def\centeron#1#2{{\setbox0=\hbox{#1}\setbox1=\hbox{#2}\ifdim
\wd1>\wd0\kern.5\wd1\kern-.5\wd0\fi
\copy0\kern-.5\wd0\kern-.5\wd1\copy1\ifdim\wd0>\wd1 
\kern.5\wd0\kern-.5\wd1\fi}}
\def\centerover#1#2{\centeron{#1}{\setbox0=\hbox{#1}\setbox
1=\hbox{#2}\raise\ht0\hbox{\raise\dp1\hbox{\copy1}}}}
\def\centerunder#1#2{\centeron{#1}{\setbox0=\hbox{#1}\setbox
1=\hbox{#2}\lower\dp0\hbox{\lower\ht1\hbox{\copy1}}}}
\def\st#1{\centeron{$#1$}{$/$}}
\catcode`\@=11
\def\eqnarray{\stepcounter{equation}\let\@currentlabel =\theequation 
\global\@eqnswtrue
\global\@eqcnt\z@\tabskip\mathindent\let\\=\@eqncr
\abovedisplayskip\topsep\ifvmode\advance\abovedisplayskip\partopsep\fi 
\belowdisplayskip\abovedisplayskip\belowdisplayshortskip\abovedisplayskip
\abovedisplayshortskip\abovedisplayskip$$\m@th\halign to\linewidth\bgroup
\@eqnsel\hskip\@centering
$\displaystyle\tabskip\z@{##}$&\global\@eqcnt\@ne 
\hfil${##}$\hfil 
&\global\@eqcnt\tw@ 
$\displaystyle{##}$\hfil
\tabskip\@centering&\llap{##}\tabskip\z@ \cr}
\catcode`\@=12
\newcommand{\AmS}{{\protect\the\textfont2
  A\kern-.1667em\lower.5ex\hbox{M}\kern-.125emS}}

\title{\vskip -3.55cm\rightline{
\vbox{\normalsize
\halign{&## \hfil\cr
&ANL-HEP-CP-96-57\cr
&August, 1996\cr}
}}\vskip 2.10cm%
A lattice formulation of chiral gauge theories%
\thanks{Talk presented at the Lattice~'96 Symposium, St.~Louis, Missouri,
June, 1996.}}

\author{Geoffrey T.~Bodwin\address{Argonne National Laboratory, \\ 
        9700 S.~Cass Ave., Argonne, IL 60439, USA}}
       
\begin{document}

\begin{abstract}
We present a method for implementing gauge theories of chiral fermions 
on the lattice.
\end{abstract}

\maketitle

A more detailed account of many of the ideas presented here and a more 
complete set of references can be found in Ref.~\cite{bodwin}.

\section{THE LATTICE AS A UV REGULATOR}

The essential criterion for evaluating the suitability of a lattice
formulation of a field theory is the extent to which the lattice
fulfills the r\^ole of a UV regulator.  A good UV regulator has the
following properties. (1)~The noninteracting theory yields the correct
spectrum as the regulator is removed ($a\rightarrow 0$). (2)~The
interacting theory yields the correct action as the regulator is
removed, holding field momenta fixed (and neglecting irrelevant
operators). (3)~The regulated theory respects the ``important''
symmetries. 

A theory that exhibits property~(2) respects the symmetries of the
unregulated (continuum) theory as the regulator is removed, holding
field momenta fixed. However, if momenta of order the cutoff are
important, these symmetries can be violated even once the regulator has
been removed. The symmetry violations arise from UV-divergent
amplitudes, and, hence, they correspond to short-distance effective
interactions (local renormalization counterterms). 

There are many important symmetries. However, in discussing gauge
theories, we focus on the gauge symmetry because it is a powerful tool
to proscribe symmetry-violating counterterms. The gauge symmetry
generally allows only those counterterms, such as coupling-constant
renormalization, that are compatible with the symmetries of the
continuum theory. Note that invariance under {\it infinitesimal} gauge
transformations is usually sufficient to rule out gauge-variant
counterterms.

\section{A MODEL: CHIRAL QED}

As a concrete example of a lattice chiral gauge theory, let us examine
four-dimensional QED with left-handed couplings to the gauge field. We
take the standard plaquette action for the gauge field, and we begin
with the simplest (``naive'') discretization of the gauged fermion
action: 
\begin{eqnarray}
S_{N}=a^4\sum_{x,\mu}&&\overline\psi(x)\gamma_\mu P_L
{1\over 2a}
[U_\mu(x)\psi(x+a_\mu)\nonumber\\
&&\qquad-U_\mu^\dagger(x-a_\mu)\psi(x-a_\mu)],
\end{eqnarray}
where $P_{R/L}=(1/2)(1\pm \gamma_5)$, and, as usual,
\begin{equation}
U_\mu(x)\equiv \exp[iagA_\mu(x+a_\mu/2)].
\label{link}
\end{equation}

This action has the well-known doubling problem, which violates
property~(1): the fermion propagator
\begin{equation}
iS_F^N=[(1/a)\sum_\mu i\gamma_\mu\sin(p_\mu a)]^{-1}
\end{equation}
has poles at $p_\mu=0$ and $p_\mu=\pi/a$.
The doublers can be eliminated in the usual way by introducing a Wilson
mass term 
\begin{eqnarray}
S_{WI}=a^4\sum_{x,\mu}&&\overline\psi(x) {1\over 2a}
[2\psi(x)-U_\mu(x)\psi(x+a_\mu)\nonumber\\
&&\qquad-U_\mu^\dagger(x-a_\mu)\psi(x-a_\mu)].
\label{wilson}
\end{eqnarray}
Now the propagator is 
\begin{equation}
iS_F^W=(1/a)\sum_\mu [i\gamma_\mu \sin(p_\mu a)+M(p)]^{-1},
\end{equation}
where $M(p)=(1/a)\sum_\mu[1-\cos(p_\mu a)]$ is the Wilson mass. There is
a pole only at $p_\mu=0$, and the lattice action satisfies properties~1
and 2. 

However, the action does not satisfy property~(3).
The Wilson mass term violates the chiral symmetry, since it
connects $P_L\psi$ to $\overline \psi P_R$.  Consequently, the action is
not invariant under an infinitesimal gauge transformation 
\begin{eqnarray}
&&U_\mu(x)\rightarrow U_\mu(x)\!+\!i\Lambda(x)U_\mu(x)
\!-\!iU_\mu(x)\Lambda(x\!+\!a_\mu),\nonumber\\
&&\psi(x)\rightarrow [1+iP_L\Lambda(x)]\psi(x),
\end{eqnarray}
which involves only the left-handed fermion field. 
The gauge variation of the action is nonzero because the Wilson mass
operator commutes, rather than anticommutes, with $\gamma_5$. 

The gauge variation gives a $\Lambda$-fermion vertex, which, in momentum 
space, is
\begin{equation}
{\cal M}=i(P_R-1)M(p)-i(P_L-1)M(p+l),
\end{equation}
where $p$ is the fermion momentum, and $l$ is the
gauge-field momentum. Now, $M(p)$ vanishes as $ap^2$ in the limit
$a\rightarrow 0$ with $p$ fixed. Therefore, gauge  variations can
survive in the continuum limit only if $p$ or $l$ of order $1/a$ are
important, {\it i.e.}, only in UV-divergent expressions. 

\section{MODIFICATION OF THE FERMION DETERMINANT}

We can try to restore the gauge symmetry by modifying fermion
amplitudes.  Since the difficulties arise from the fact that $\gamma_5$
commutes with the Wilson term, let us adopt the strategy of treating
$\gamma_5$ as if it {\it anticommuted} with the Wilson term.  Of course,
we make an error in doing this.  However, the error is proportional to
Wilson terms, which vanish as $a\rightarrow 0$, unless $p\sim 1/a$.
Therefore, the procedure amounts to adding local interactions
(counterterms) to the theory, {\it i.e.}, to modifying the UV regulator.

Consider a typical fermion-loop amplitude:
\begin{equation}
{\rm Tr}\, \ldots (\st p_1+M)^{-1}\Gamma_\mu P_L
(\st p_2+M)^{-1}\Gamma_\nu P_L\ldots,
\end{equation}
where the $\Gamma$'s are gauge-field-fermion vertices and we are using a 
schematic notation for the fermion propagators. For even-parity
terms, (those containing an even number of $\gamma_5$'s), we treat
$\gamma_5$ as if it commuted with all propagators and vertices, {\it
including Wilson terms}, and use the identity $\gamma_5^2=1$ to eliminate 
$\gamma_5$'s. The even-parity part of the amplitude is then
\begin{equation}
(1/2)\,{\rm Tr}\, \ldots(\st p_1+M)^{-1}\Gamma_\mu(\st
p_2+M)^{-1}\Gamma_\nu\ldots.
\end{equation}
This is $1/2$ the amplitude for a Dirac particle with vector-like
coupling to the gauge field.  Now, the vector-like amplitude {\it is}
gauge invariant, provided that we gauge the Wilson term.  (In this case
we must transform $\psi$, not just $P_L\psi$.) Since $\det {\cal
D}=\exp( {\rm Tr}\ln {\cal D})$, where ${\rm Tr}\ln {\cal D}$ is the
amplitude, $1/2$ the amplitude corresponds to the square root
of the fermion determinant. It is easy to see that the even-parity part
of the amplitude yields the magnitude of the determinant. Therefore, we
can render the magnitude of the determinant {\it exactly} gauge
invariant by replacing it with the square root of the determinant for a
Wilson-Dirac particle with vector-like coupling to the gauge field.
(There is a similar modification for interacting fermion propagators 
\cite{bodwin}.)

In the case of the odd-parity terms, this trick doesn't work.  There is
always one $\gamma_5$ that can't be eliminated. However, the Dirac trace
of an odd number of $\gamma_5$'s is proportional to
$\epsilon_{\mu\nu\rho\sigma}$.  Therefore, as $a\rightarrow 0$ with
gauge-field momenta fixed, the gauge variation is proportional to
operators with dimension less than five containing
$\epsilon_{\mu\nu\rho\sigma}$. These are precisely the operators that
correspond to the ABJ anomaly. Thus, anomaly species cancellation
guarantees the gauge symmetry for the odd-parity loop amplitudes---as
long as we consider only gauge-field configurations with momenta much
less than $\pi/a$. 

\section{LARGE GAUGE-FIELD MOMENTA}

In odd-parity contributions, gauge-field momenta $l \sim 1/a$ flowing
through a fermion line yield nonvanishing gauge variations, since
$M(p+l)$ is not negligible. Gauge-field momenta $l \sim 1/a$ are
important in UV-divergent loops involving gauge fields. Such loops
generate all local interactions that are consistent with the symmetries
of the lattice theory.  Many of the interactions are gauge variant and
must be eliminated---possibly by tuning of the corresponding
counterterms. 

Let us now discuss a method to avoid explicit tuning of counterterms.
The essential idea is to introduce {\it separate} lattice spacings
(cutoffs) $a_g$ for the gauge field and $a_f$ for the fermion field.
Then $a_f$ is taken to zero before $a_g$.  The goal of such a procedure
is to suppress amplitudes in which gauge-field momenta are greater in
magnitude than $\pi/a_g$. Then Wilson terms in propagator numerators and
vertices can be neglected, except in divergent fermion loops, which we
have already dealt with by modifying the fermion determinant and
invoking anomaly cancellation. 

This idea is implemented as follows. One treats links on the $a_g$
lattice as the dynamical variables: the gauge-field action depends only
on these links.  The fermion-field action---including the interactions
with the gauge field---resides on the $a_f$ lattice.  One obtains the
links on the $a_f$ lattice by interpolating the gauge fields on the
$a_g$ lattice. 

A good interpolation must have certain properties: (1)~gauge covariance;
(2)~continuity on the $a_f$ lattice, except at boundaries of $a_g$
plaquettes, (3)~locality. Property~(1) means that a gauge transformation
on the $a_g$ lattice maps into a gauge transformation on the $a_f$
lattice. Property~(2) guarantees that the Fourier transform of the gauge
field is suppressed as $1/(a_gl)$ for gauge-field momenta $l\gg 1/a_g$.
Property~(3) insures that one recovers the continuum theory for $la_g\ll
1$. Under these assumptions, one can prove, to all orders in
perturbation theory, that, as $a_f\rightarrow 0$ with $a_g$ fixed, gauge
variations vanish as $a_f$, and deviations from the target theory vanish
as $a_f^2$ (Ref.~\cite{bodwin,hernandez-sundrum}). This result also
holds in the presence of individual nonperturbative gauge-field
configurations \cite{bodwin}. 

We assume that the lattice origins coincide and that $a_g/a_f$ is an
integer.  Along $a_g$-lattice links $U_g$, we take the gauge-field on
the $a_f$ lattice, $A_f$, to be 
\begin{equation}
A_{f\mu}=(1/ia_g g)\log U_{g\mu},
\label{plaquette-interp}
\end{equation}
which is consistent with gauge covariance.  Away from the $a_g$ links,
one obtains $A_{f\mu}$ by interpolation, using (\ref{plaquette-interp})
as a boundary condition. For Abelian gauge groups, a simple linear
interpolation is consistent with the properties of a good regulator.
(For non-Abelian groups the interpolation is more complicated
\cite{hernandez-sundrum}). The links on the $a_f$ lattice are then
obtained from $A_f$.

\section{A NONPERTURBATIVE PROBLEM}

Shamir has pointed out a potential difficulty with interpolation of the
gauge field \cite{shamir}. A gauge transformation on the $a_g$ lattice
$U_\mu(x)\rightarrow V^{-1}(x+a_{g\mu})U_\mu(x)V(x)$ can take $U$
through $-1$ when either the transformation is large or U is near $-1$.
Then, $gA_{f\mu}$ (Eq.~\ref{plaquette-interp}) jumps from $\pi/a$ to
$-\pi/a$. The jump corresponds to a change in the winding number.
Hence, the corresponding gauge transformation on the interpolated
field is singular. Therefore, we are faced with two alternatives: require
gauge covariance and accept that the interpolated field may be singular,
or require smoothness of the interpolated field and accept that gauge
covariance is violated for some gauge transformations. 

Let us adopt the second alternative. (In the Abelian case, we simply
take the linear interpolation of (\ref{plaquette-interp}).) Then we
still have invariance under infinitesimal gauge transformations, except
at $gA_\mu=\pi/a$.  Recall that we require only invariance under
infinitesimal transformations to prohibit undesirable counterterms.  The
issue then is whether the point $gA_\mu=\pi/a$ important in the path
integral. 

To address this question, let us consider, for example, the situation in
the gauge $\nabla_\mu A_\mu=0$. In this gauge $\nabla^2 A_\mu=J_\mu$, so
$A_\mu\sim 1/a$ can arise only from currents (momenta) $\sim 1/a$.  The
most divergent amplitude is the vacuum polarization. By dimensional
analysis the vacuum polarization $\Pi_{\mu\nu}$ goes as $\delta_{\mu\nu}
(1/a_g^2)$.  However, invariance under infinitesimal gauge
transformations reduces this to $\Pi_{\mu\nu}\sim
(k^2\delta_{\mu\nu}-k_\mu k_\nu) \log (1/a_g)$. This result can be
verified by explicit calculation in the Abelian theory with the linear
interpolation. Therefore, $A_\mu$ grows only as $\log(1/a)$. We conclude
that the point at $gA_\mu=\pi/a$ doesn't contribute significantly to the
path integral as $a\rightarrow 0$.

\end{document}